\documentclass[%
 reprint,
 amsmath,amssymb,
 aps,
pra,
]{revtex4-2}

\usepackage{amsmath,amssymb}
\usepackage{bm}
\usepackage{graphicx}
\usepackage{dcolumn}
\usepackage{tabularx}
\usepackage{enumitem}
\usepackage{threeparttable}
\usepackage{booktabs}
\usepackage{cancel}
\usepackage[normalem]{ulem}
\usepackage{xcolor}
\usepackage{physics}
\usepackage{float}

\graphicspath{{images/}}
\usepackage{orcidlink}
\usepackage{hyperref}
\hypersetup{
    colorlinks=true,
    citecolor=blue,
    linkcolor=blue,
    urlcolor=blue
}
\begin{document}

\title{Influence of $\Delta$ resonances in hyperonic neutron stars on gravitational-wave inspiral signatures}

\author{Athira S.\orcidlink{0009-0005-3818-7186}$^{1}$}
\email{p23ph0002@iitj.ac.in}
\author{Vivek Baruah Thapa\orcidlink{0000-0003-2411-9885}$^{2}$}
\email{vivekthapa@bssrv.ac.in}
\author{Monika Sinha\orcidlink{0000-0002-3080-9949}$^{1}$}
\thanks{Corresponding author}
\email{ms@iitj.ac.in}

\affiliation{$^{1}$ Indian Institute of Technology Jodhpur, Jodhpur 342030, India}
\affiliation{$^{2}$ Birangana Sati Sadhani Rajyik Vishwavidyalaya, Assam 785621, India}

\begin{abstract}
The composition of dense matter in neutron star interiors remains uncertain, with the possible appearance of the non-strange $\Delta$ resonances providing an important source of theoretical uncertainty. In this work, we investigate the gravitational wave signatures of hyperonic neutron stars containing $\Delta$ baryons within a density-dependent relativistic mean-field framework based on the DDME2 parametrization. Within this framework, the baryons interact among themselves via mesons. For $\Delta$ baryons, the interaction is uncertain to date. We investigate how variations of the scalar and vector $\Delta-$meson couplings, subject to \(x_{\sigma\Delta}-x_{\omega\Delta}=0.2\), modify the tidal properties and inspiral gravitational-wave signals.
We find that the allowed variation in the $\Delta-\sigma$ coupling leads to substantial changes in the tidal properties of the neutron stars, which in turn leave distinct imprints on the gravitational-wave waveforms, particularly in their frequency evolution, accumulated phase, and strain amplitude. \textcolor{black}{These waveform differences arise through changes in tidal deformability and are not unique signatures of $\Delta$ resonances. Future gravitational-wave observations may therefore constrain the $\Delta-$meson couplings within the adopted dense-matter framework, subject to degeneracies with alternative EOS models.}
\end{abstract}

\maketitle

\section{Introduction}
Neutron stars (NSs) provide a unique laboratory for exploring strongly interacting matter under extreme conditions that are inaccessible to terrestrial experiments. Recent observations of massive NSs suggest that the densities attained in their interiors may be sufficiently high for heavier baryonic degrees of freedom to appear in addition to nucleons \cite{2023PrPNP.13104041S, 2024arXiv240214288W, 2023PhRvC.107d5804M, 2024AcPPB..55....1T}. At supranuclear densities, the composition of matter is therefore expected to include strange baryons (hyperons) as well as nonstrange baryons such as the $\Delta$ resonances \cite{2007PrPNP..59...94W, 2011PhRvC..83b5805W, 2023PhRvC.107d5804M}. Among these, $\Delta$ resonances have attracted increasing attention because they may appear at relatively low densities, depending on the strength of their in-medium interactions \cite{2024arXiv241201201J, 2021EPJA...57..216D}. Among the various theoretical models developed to describe dense matter, the relativistic mean-field (RMF) model is one of the most widely used frameworks \cite{1974AnPhy..83..491W, Parmar_2021, 2024PhRvD.110k4009X}. In this approach, baryons interact through the exchange of mesons. However, the couplings of the $\Delta$ baryons to the meson fields remain poorly constrained. The relevant mesons include the isoscalar-scalar $\sigma$ meson, the isoscalar-vector $\omega$ and $\phi$ mesons, and the isovector-vector $\rho$ meson \cite{1995PhRvC..52.3470K, 1985ApJ...293..470G}. Consequently, comparing the properties of dense matter predicted by the RMF model with observations of  NSs provides an important means of constraining these coupling strengths and testing the validity of the model.

The advent of gravitational-wave (GW) astronomy has opened a new observational window into the internal structure of NSs \cite{Huang_2026}. Measurements of the early part of the gravitational wave signal from binary neutron star mergers, such as those reported by the LIGO \cite{LIGOwebsite}--Virgo \cite{VirgoWebsite}--KAGRA \cite{KAGRAWebsite} collaboration, encode information about the tidal deformability and compactness of NSs, which are directly sensitive to the underlying dense matter physics \cite{PhysRevD.81.123016, PhysRevD.93.064082}. In Ref. \cite{PhysRevD.88.044042}, hybrid waveforms were constructed for the first time by combining numerical-relativity waveforms with resummed post-Newtonian (PN) waveforms based on the Taylor-T4 approximant, and the prospects for measuring tidal deformability from GW observations were investigated. Since tidal deformability depends sensitively on the NS equation of state (EOS), comparisons between theoretical GW waveforms and observations can be used to constrain the composition of dense matter, including the possible appearance of $\Delta$ resonances \cite{2018PhRvC..98c5804M, 2019PhRvD..99b3004S}. 
The modifications of the EOS induced by $\Delta$ degrees of freedom may, in principle, be reflected in tidal observables and the late-inspiral GW phasing. Establishing whether these modifications are sufficiently large to constrain the underlying $\Delta$-meson interaction requires quantitative assessment of their effects on the inspiral dynamics.
Specifically, the poorly constrained $\Delta$ meson coupling strengths, often parameterized through the $\Delta$--$\sigma$ potential, play a crucial role in determining the onset density and population of $\Delta$ baryons, thereby influencing the stiffness of the EOS and the global properties of NSs.

In this work, we investigate the role of $\Delta$ baryons in shaping GW observables and examine how variations in the $\Delta$--$\sigma$ potential influence both the EOS and the resulting GW signal. While the effects of $\Delta$ baryons on the EOS and the bulk properties of NSs have been investigated in several previous studies \cite{2010ApJ...724L..74S, 2020PhLB..80235266M, 2020PhRvD.102f3008M, 2025PhRvD.112b3016P, 2022EPJA...58..137S}, their imprint on the GW waveform remains comparatively unexplored. Since the GW strain, phase, and frequency encode the dynamical evolution of compact binary mergers, they provide a direct observational probe of the underlying EOS \cite{2015PhRvD..92j4008C, 2009JPhCS.189a2024M, 2009APS..APRR11004M, 2015APS..APR.C3001L, 2014APS..APRB15001L}. Motivated by these developments, we systematically vary the $\Delta$--$\sigma$ potential within a relativistic mean-field framework and investigate how the resulting modifications to the EOS are reflected in the evolution of the GW strain, phase, and frequency as functions of the retarded time, using the post-Newtonian (PN) formalism with the T4 approximant \cite{2012PhRvD..85j4030B, 2009PhRvD..79l4033R, 2007PhRvD..76l4038B}. We also compute the corresponding mass--radius relations and tidal deformabilities and compare them with current GW observations and constraints. Our objective is to quantify the influence of $\Delta$ baryons on neutron-star structure and to assess whether GW waveform observables can provide meaningful constraints on the poorly known $\Delta$--meson interaction.

\textcolor{black}{Our analysis quantifies how variations of the meson–\(\Delta\) couplings affect stellar tidal properties and the associated inspiral waveforms within a fixed DDME2 framework. It does not establish a composition-specific GW signature. In the adopted adiabatic quadrupolar tidal approximation, the internal structure enters the inspiral dynamics through the component tidal deformabilities. Consequently, different microscopic compositions can produce indistinguishable waveforms when their masses and tidal deformabilities coincide. Any inference of the \(\Delta\)-meson couplings from these observables is therefore conditional on the assumed nucleonic interaction, hyperon couplings, and coupling relations. We examine this conditional sensitivity through a one-dimensional coupling scan and waveform comparisons performed over a common frequency interval.}
The remainder of this paper is organized as follows. In Sec.~\ref{formalism}, we present the theoretical framework adopted to describe dense NS matter and the formalism used to calculate its macroscopic and GW properties. The results of our analysis and their astrophysical implications are discussed in Sec.~\ref{results}. Finally, our conclusions are summarized in Sec.~\ref{Conclusions}.

\section{Formalism}
\label{formalism}
\subsection{RMF Model}
The dense matter inside the NS core is described within the framework of a density-dependent relativistic mean-field (DD-RMF) model under the assumptions of charge neutrality and $\beta$ equilibrium. At low density, the matter is composed of nucleons ($n,p$), leptons - electrons ($e^-$) and muons ($\mu^-$), which ensure charge neutrality at all densities. In addition to the nucleonic degrees of freedom, with the increase of density we consider the appearance of the hyperons ($\Lambda, \Sigma$ and $\Xi$) and the spin-$3/2$ $\Delta$-baryon quartet, namely $\Delta^{++}$, $\Delta^{+}$, $\Delta^{0}$, and $\Delta^{-}$, whose population at supranuclear densities can significantly influence the composition and EOS of dense matter. The baryons interact through the exchange of the isoscalar-scalar ($\sigma$), isoscalar-vector ($\omega$), isovector-vector ($\rho$), and hidden-strangeness vector ($\phi$) mesons, with density-dependent meson-baryon coupling constants that effectively account for many-body correlations in the nuclear medium. The density dependence of the couplings enables the model to reproduce the empirical properties of finite nuclei and nuclear matter while providing a realistic description of matter at the extreme densities encountered in NS interiors. Within the RMF formalism, the interactions among baryons mediated by meson exchange are characterized by the following total Lagrangian density \cite{LI2018234, 10.1093/mnras/stab2327}:

\begin{align}
\mathcal{L}=&\sum_{b}\bar{\psi}_b\left(i\gamma_\mu D_b^\mu-m_b^*\right)\psi_b+\sum_l\bar{\psi}_l\left(i\gamma_\mu \partial^\mu-m_l\right)\psi_l\nonumber\\&+\sum_{\Delta}
\bar{\psi}_{\Delta\nu}\left(i\gamma_\mu D_\Delta^\mu-m_\Delta^*\right)\psi_\Delta^\nu\nonumber\\&
+\frac{1}{2}\left(\partial_\mu\sigma \partial^\mu\sigma-m_\sigma^2\sigma^2\right)-\frac{1}{4}
\omega_{\mu\nu}\omega^{\mu\nu}+\frac{1}{2}m_\omega^2\omega_\mu\omega^\mu\nonumber\\&
-\frac{1}{4}\bm{\rho}_{\mu\nu}\cdot\bm{\rho}^{\mu\nu}+\frac{1}{2}m_\rho^2\bm{\rho}_\mu\cdot\bm{\rho}^\mu-\frac{1}{4}
\phi_{\mu\nu}\phi^{\mu\nu}+\frac{1}{2}m_\phi^2\phi_\mu\phi^\mu.
\end{align}
Here, the fields $\psi_b$, $\psi_l$, and $\psi_{\Delta}^{\nu}$ represent the
Dirac spinors corresponding to the baryon octet ($b= n, p, \Lambda, \Sigma, \Xi$), leptons ($e^-, \mu^-$), and the
Rarita--Schwinger spin-$3/2$ fields describing the $\Delta$-baryon
quartet ($\Delta^{++}$, $\Delta^{+}$, $\Delta^{0}$, $\Delta^{-}$), respectively. The quantities $m_b$, $m_l$, and $m_{\Delta}$
denote the corresponding bare masses of the baryons, leptons, and
$\Delta$ resonances. The interaction of these particles with the
vector meson fields is incorporated through the covariant derivative,
which is defined as \cite{2001PhRvC..64e5805B}

\begin{equation}
D_{\mu}^{(j)}
=
\partial_{\mu}
+i g_{\omega j}\omega_{\mu}
+i g_{\rho j}\bm\tau_{3j}\cdot\bm\rho_{\mu}
+i g_{\phi j}\phi_{\mu},
\end{equation}
where $j$ labels the baryon species, i.e., the nucleons, hyperons, and
$\Delta$ baryons, and $\bm\tau_{3j}$ is the third
component of its isospin. 
$\bm\tau_j$ represents the isospin operator for the isovector-vector meson fields. The gauge field strength tensors for the vector meson fields are provided by \cite{2021PhRvD.103f3004T}
\begin{eqnarray}
 \begin{aligned}   
\omega_{\mu \nu} = \partial_{\mu}\omega_{\nu} - \partial_{\nu}\omega_{\mu}, \\
\phi_{\mu \nu} = \partial_{\mu}\phi_{\nu} - \partial_{\nu}\phi_{\mu}, \\
\boldsymbol{\rho}_{\mu \nu} = \partial_{\mu} \boldsymbol{\rho}_{\nu} - \partial_{\nu}\boldsymbol{\rho}_{\mu}, \\
\end{aligned}
\end{eqnarray}

In our work, we consider the coupling to be density-dependent. In the model considered here, the isoscalar meson-nucleon couplings vary with density as
\cite{10.1093/mnras/stab2327, 2020Parti...3..660T}

\begin{equation}\label{eqn.dd_isoscalar}
g_{i N}(n)= g_{i N}(n_{0}) f_i(x), \quad \quad \text{for }i=\sigma,\omega
\end{equation}
where 
\begin{equation}\label{eqn.func}
f_i(x)= a_i \frac{1+b_i (x+d_i)^2}{1+c_i (x +d_i)^2}.
\end{equation}
Here $x=n/n_0$, where $n_0$ is the nuclear saturation density, and n is the total baryon number density. The density dependence of the coupling parameters is described by the parameters $a_i$, $b_i$, $c_i$, and $d_i$. The isovector-vector $\rho$-meson coupling is provided by the expression \cite{ 2020Parti...3..660T} 
 
\begin{equation}
g_{\rho N}(n)= g_{\rho N}(n_{0}) e^{-a_{\rho}(x-1)}.
\end{equation}

The density argument of the couplings is the total baryon number density, including nucleons, hyperons and \(\Delta\) baryons. The rearrangement contribution associated with the density dependence of the couplings is included consistently in the baryonic chemical potentials and pressure \cite{2001PhRvC..64c4314H}. At zero temperature, thermodynamic consistency is checked using \(P=\sum_i\mu_i n_i-\varepsilon\), where the sum includes all populated baryonic and leptonic species.

\subsection{Coupling Parameters}
In this work, we will be using the DD-RMF model with DDME2 parametrization for the meson-baryon couplings \cite{2020PhRvD.102l3007T}. For the nucleonic sector, the parameter values are tabulated in Table \ref{table:vector_coupling} \cite{2005PhRvC..71b4312L, 2020PhRvD.102l3007T} with nucleon mass $m_N = 938.9$ MeV.
\begin{table*}
\begin{center}
 \caption{Parameters for the DDME2 parametrization at $n_0=0.152~fm^{-3}$ and meson masses. \cite{2005PhRvC..71b4312L}}
 \resizebox{0.5\textwidth}{!}{%
\begin{tabular}{ccccccc}
     \hline
  Meson(i)  &$g_{iN}$ & $a_i$&$b_i$&$c_i$&$d_i$ & $m_i$(MeV)\\
   \hline
  $\sigma$  & 10.5396 & 1.3881 & 1.0943 &1.7057&0.4421& 550.1238 \\
  $\omega$  & 13.0189  & 1.3892 & 0.9240 & 1.4620&0.4775& 783 \\
  $\rho$  & 7.3672 & 0.5647 &  & & & 763\\
   \hline
\end{tabular}
}%
\label{table:vector_coupling}
\end{center}
\end{table*}
\noindent The strange meson $\phi$ does not couple with nucleons, so the value of $g_{\phi N }= 0$. We employ the SU(6) flavour symmetry and quark model and isospin counting rule (QMIC) \cite{1984PrPNP..12..171D, 1994AnPhy.235...35S, S__2025, 2014PhRvC..90a5801C} for the meson-hyperon vector couplings. We consider the optical potentials of $\Lambda, \Sigma$ and $\Xi$ hyperons in symmetric nuclear matter to be $-28$ \cite{1997PhR...287..385B, 1988PhRvC..38.2700M, 2017PhRvC..95b5201T, 2018JCAP...05..038L}, $+30$ \cite{2006PhRvC..74f4613K, 2005NuPhA.759..143H, 2006NuPhA.767..206H,2007PhR...452...89F, 2000PhRvC..62c4311S} and $-14$ MeV \cite{2007arXiv0704.3929M}, respectively, at nuclear saturation density for the meson-hyperon scalar couplings \cite{2020Parti...3..660T,2021PhRvD.103f3004T}.

 Owing to the limited experimental constraints on the $\Delta$-nucleon interaction, the meson--$\Delta$ coupling constants are generally treated as free parameters. Experimental studies, including pion--nucleus scattering, electron--nucleus scattering, and $\Delta$-quartet excitation measurements, have provided constraints on these couplings at the nuclear saturation density $n_0$. Existing analyses suggest that the $\Delta$-potential in nuclear matter lies within the range $ -30~\mathrm{MeV} + V_N \leq V_{\Delta} \leq V_N,$ where $V_N$ denotes the nucleon potential \cite{2021PhRvD.103f3004T}. Furthermore, the difference between the scalar and vector coupling ratios is constrained to satisfy \cite{PhysRevC.106.055801}
\begin{equation}
0 \leq x_{\sigma\Delta} - x_{\omega\Delta} \leq 0.2,
\end{equation}
where
\begin{equation}
x_{\sigma\Delta}=\frac{g_{\sigma\Delta}}{g_{\sigma N}}, \qquad
x_{\omega\Delta}=\frac{g_{\omega\Delta}}{g_{\omega N}}.
\end{equation}
\textcolor{black}{We investigate the coupling sequence \(1.23\leq x_{\sigma\Delta}\leq1.50\), imposing \(x_{\omega\Delta}=x_{\sigma\Delta}-0.2\) and fixing \(x_{\rho\Delta}=1.0\). Thus, the scalar and vector couplings vary simultaneously, and the resulting changes cannot be attributed to the scalar coupling alone. This prescription samples the boundary \(x_{\sigma\Delta}-x_{\omega\Delta}=0.2\) \cite{1989NuPhA.504..797W, 2014PhRvC..90f5809D} of the adopted coupling interval; it is a one-dimensional sensitivity study rather than an exhaustive exploration of the meson–\(\Delta\) parameter space. The sampled lower endpoint approximately corresponds to the onset of configurations supporting a maximum mass of \(2.0\,M_\odot\) within this sequence. This selection is not interpreted as a statistically inferred lower bound on \(x_{\sigma\Delta}\). Since the \(\Delta\) quartet is nonstrange, we adopt \(g_{\phi\Delta}=0\).}

\subsection{Mass-Radius relation}

The equation of state obtained within the DD-RMF framework provides
the relation between the pressure $P$ and the energy density $\varepsilon$
of dense matter. This EOS is then used to determine the macroscopic
properties of NSs by solving the Tolman--Oppenheimer--Volkoff
(TOV) equations \cite{1939PhRv...55..364T, 1939PhRv...55..374O, 1997A&A...328..274B} under the assumption of hydrostatic equilibrium and
spherical symmetry. In geometrized units ($G=c=1$), the TOV equations are
given by
\begin{equation}
\frac{dP}{dr}
=
-\frac{[\varepsilon(r)+P(r)]
[M(r)+4\pi r^3P(r)]}
{r[r-2M(r)]},
\label{eq:TOV_pressure}
\end{equation}
and
\begin{equation}
\frac{dM}{dr}
=
4\pi r^2\varepsilon(r),
\label{eq:TOV_mass}
\end{equation}
where $M(r)$ is the gravitational mass enclosed within a radius $r$,
$P(r)$ is the pressure, and $\varepsilon(r)$ is the corresponding energy
density. For a given central energy density $\varepsilon_c$, the
equations are integrated outward from the stellar centre, with the
boundary conditions
\begin{equation}
M(0)=0,
\qquad
P(0)=P_c,
\end{equation}
until the pressure decreases to zero. The corresponding radius defines
the stellar radius $R$, and the enclosed mass at this point gives the
gravitational mass,
\begin{equation}
M=M(R).
\end{equation}

By varying the central density, a sequence of equilibrium NS configurations is obtained, yielding the mass--radius (M-R) relation, with the Baym–Pethick–Sutherland (BPS) EOS used to model the crust \cite{1971ApJ...170..299B}.
For each EOS, we check mechanical stability and causality over the density range sampled by the stable stellar configurations. Specifically, we verify \(dP/d\varepsilon>0\) and \(c_s^2/c^2=dP/d\varepsilon\leq1\). Stellar models used in the comparisons are selected from the stable branch of the mass–central-density sequence.
The maximum mass supported by the EOS corresponds to the maximum of this sequence and provides an important constraint on the stiffness of dense matter. In the present work, the appearance of hyperons and $\Delta$ baryons modifies the particle composition and consequently the pressure and energy density of the stellar matter. In particular, variations in the scalar $\Delta$-meson coupling $x_{\sigma\Delta}$, together with the
corresponding variation of $x_{\omega\Delta}$ subject to the constraint on $x_{\sigma\Delta}-x_{\omega\Delta}$, modify the $\Delta$ population and the resulting EOS. The resulting changes in the EOS are therefore reflected in the NS M-R relation.

\subsection{Tidal Deformability}

The tidal deformability quantifies the response of a NS to the external tidal field produced by its binary companion during the inspiral phase. It is defined through the relation between the induced mass quadrupole moment, $Q_{ij}$, and the external static tidal field, $E_{ij}$ \cite{Hinderer_2008, PhysRevD.81.123016, PhysRevD.99.023009, Glendenning:1997wn}, as
\begin{equation}
Q_{ij}=-\lambda E_{ij},
\end{equation}
where $\lambda$ is the tidal deformability parameter. It is related to the dimensionless quadrupolar Love number, $k_2$, through 
\begin{equation}
\Lambda=\frac{\lambda}{M^5}=\frac{2}{3}k_2C^{-5},
\label{lambda}
\end{equation}
where $\Lambda$ is the dimensionless tidal deformability parameter and $C=M/R$ is the compactness of the NS (in geometrized units, $G=c=1$). The dimensionless Love number $k_2$ is obtained by solving the perturbation equations together with the TOV equations and is given by \cite{Hinderer_2008, PhysRevD.80.084035}
\begin{equation}
\begin{aligned}
k_2=&\,
\frac{8C^5}{5}(1-2C)^2
\left[2+2C(y_R-1)-y_R\right]
\\
&\times
\Bigg\{
2C\Big[6-3y_R+3C(5y_R-8)\Big]
\\&
+4C^3\Big[13-11y_R+C(3y_R-2)
\\&
+2C^2(1+y_R)\Big]
+3(1-2C)^2
\Big[2-y_R
\\&
2C(y_R-1)\Big]
\log(1-2C)
\Bigg\}^{-1},
\end{aligned}
\label{eq:k2}
\end{equation}
where $y_R=y(R)$ is the value of the metric perturbation function evaluated at the stellar surface.

For the canonical configurations, the stellar sequence is interpolated to a gravitational mass of exactly \(1.4\,M_\odot\). The radius, Love number, compactness, and tidal deformability are evaluated for this same configuration. We verify the identities \(C=GM/(Rc^2)\) and \(\Lambda=(2/3)k_2C^{-5}\) before using these quantities in the waveform calculations.

\subsection{Post-Newtonian GW Formalism}

The gravitational-wave signal from a binary neutron-star (BNS) coalescence can be broadly divided into three stages: inspiral, merger, and ringdown \cite{Baiotti_2017, 2021GReGr..53...27D}. During the inspiral, the binary loses orbital energy and angular momentum through GW emission, causing the orbital separation to decrease while the GW frequency and amplitude increase, producing the characteristic chirp signal \cite{2016NuPhA.956..225S, 2020ARNPS..70...95R}. The inspiral can be described within the PN approximation when the orbital velocity remains sufficiently smaller than the speed of light \cite{Blanchet_2002,2014arXiv1410.7832B, Das_2021}. As the stars approach each other, finite-size and tidal effects become increasingly important, eventually leading to the highly nonlinear merger phase, which generally requires numerical relativity calculations. The post-merger remnant may subsequently undergo ringdown or collapse into a black hole, depending on the total mass and the EOS \cite{2013PhRvL.111m1101B, 2017MNRAS.471.4956B, 2020PhRvL.125n1103B}.

In the present work, we focus on the late-inspiral phase, where tidal interactions become significant, and the GW signal becomes sensitive to the internal structure of the NSs \cite{2012PhRvD..85l3007D, 2010PhRvD..81h4016D}. The tidal response is characterized by the dimensionless tidal deformability given in Eq. \eqref{lambda}.

To model the inspiral waveform, we employ the Taylor-T4 PN approximant \cite{PhysRevD.93.064082, 2024PhLB..85839076H}, which is based on the energy-balance relation
\begin{equation}
\mathcal{L}=-\frac{dE}{dt},
\end{equation}
where $E$ is the orbital binding energy and $\mathcal{L}$ is the GW luminosity.\\ Introducing the gauge-independent PN parameter
\begin{equation}
x=(M_{\rm tot}\Omega)^{2/3}
=(\pi M_{\rm tot}f)^{2/3},
\end{equation}
where $M_{\rm tot}=M_1+M_2$ is the total mass of the binary, $\Omega$ is the orbital angular velocity, and $f$ is the GW frequency; the orbital evolution is obtained from
\begin{equation}
\frac{dx}{dt}
=
-\frac{\mathcal{L}(x)}
{dE(x)/dx}.
\label{eq:t4_balance}
\end{equation}

For a quasi-circular binary, the orbital binding energy and GW luminosity are expanded as a PN series in the parameter $x$. In the 3 PN approximation, the orbital energy can be written as
\begin{equation}
\begin{aligned}
E(x)
=&-\frac{M_{\rm tot}\eta x}{2}
\Bigg[
1
+\left(-\frac{3}{4}-\frac{\eta}{12}\right)x\\
&+\left(-\frac{27}{8}+\frac{19}{8}\eta
-\frac{\eta^2}{24}\right)x^2
\\&+\Bigg(
-\frac{675}{64}
+\frac{34445}{576}\eta
-\frac{205\pi^2}{96}\eta\\
&\qquad
-\frac{155}{96}\eta^2
-\frac{35}{5184}\eta^3
\Bigg)x^3
\Bigg],
\end{aligned}
\label{eq:PN_energy}
\end{equation}
where
\begin{equation}
\eta=\frac{M_1M_2}{M_{\rm tot}^2}
\end{equation}
is the symmetric mass ratio. The corresponding GW luminosity is expanded as
\begin{equation}
\begin{aligned}
\mathcal{L}
={}&
\frac{32}{5}\eta^2x^5
\Bigg\{
1
+\left(
-\frac{1247}{336}
-\frac{35}{12}\eta
\right)x
+4\pi x^{3/2}
\\
&+\left(
-\frac{44711}{9072}
+\frac{9271}{504}\eta
+\frac{65}{18}\eta^2
\right)x^2
\\
&+\left(
-\frac{8191}{672}
-\frac{583}{24}\eta
\right)\pi x^{5/2}
\\
&+\left[
\frac{6643739519}{69854400}
+\frac{16\pi^2}{3}
-\frac{1712}{105}\gamma_E
-\frac{856}{105}\ln(16x)
\right.
\\
&
\left.
+\left(
-\frac{134543}{7776}
+\frac{41\pi^2}{48}
\right)\eta
-\frac{94403}{3024}\eta^2
-\frac{775}{324}\eta^3
\right]x^3
\\
&+\left(
-\frac{16285}{504}
+\frac{214745}{1728}\eta
+\frac{193385}{3024}\eta^2
\right)\pi x^{7/2}
\Bigg\}.
\end{aligned}
\label{eq:PN_flux}
\end{equation}
where the higher-order terms are retained up to the 3.5 PN order. Substitution of the PN expansions of $E(x)$ and $\mathcal{L}(x)$ into Eq.~\eqref{eq:t4_balance}, followed by a PN re-expansion, yields the Taylor-T4 evolution equation for $x(t)$. 

The wave amplitude is decomposed into spin-weighted spherical harmonics as \cite{PhysRevD.106.124018}
\begin{equation}
h(t,\theta,\varphi)
=
\sum_{\ell,m}
h_{\ell m}(t)\,
{}_{-2}Y_{\ell m}(\theta,\varphi).
\end{equation}
where $h_{\ell m}$ are the corresponding mode amplitudes and
${}_{-2}Y_{\ell m}$ are the spin-weighted spherical harmonics. For a
quasi-circular binary, the $(2,2)$ mode provides the dominant
contribution during the inspiral. 
Retaining the displayed point-particle amplitude corrections through relative 2.5PN order, the dominant mode is expressed as
\begin{equation}
\begin{aligned}
h_{22}
={}&
-8\sqrt{\frac{\pi}{5}}\,
\frac{M_{\rm tot}\eta}{D}\,
x\,e^{-2i\Phi}
\Bigg[
1+
\left(
-\frac{107}{42}
+\frac{55}{42}\eta
\right)x\\
&
+2\pi x^{3/2}
+\left(
-\frac{2173}{1512}
-\frac{1069}{216}\eta
+\frac{2047}{1512}\eta^2
\right)x^2
\\
&+
\left(
-\frac{107\pi}{21}
+\left(\frac{34\pi}{21}-24i\right)\eta
\right)x^{5/2}
+\mathcal{O}(x^3)
\Bigg].
\end{aligned}
\label{eq:h22}
\end{equation}
where $D$ is the observer-source distance. 
No explicit tidal amplitude correction is included in this expression. At fixed component masses, distance, and GW frequency, the mode-amplitude envelope is therefore identical for the different EOSs considered here. EOS dependence enters the time-domain waveform through the frequency and phase evolution.
A representative distance of $D=100~\mathrm{Mpc}$ is adopted in our analysis. The two independent polarizations of the gravitational wave amplitude, namely the plus ($h_+$) and cross ($h_\times$)
polarizations, can be combined into the complex strain \cite{2011CQGra..28s5015R}
\begin{equation}
h(t,\theta,\varphi)
=
h_+(t,\theta,\varphi)
-i h_\times(t,\theta,\varphi).
\end{equation}
The plus and cross polarizations are subsequently obtained from the
complex strain through
\begin{equation}
h_+
=
\mathrm{Re}(h),
\qquad
h_\times
=
-\mathrm{Im}(h).
\end{equation}

The orbital phase is subsequently obtained from
\begin{equation}
\frac{d\Phi}{dt}=
\frac{x^{3/2}}{M_{\rm tot}}.
\label{eq:orbital_phase}
\end{equation}
For the dominant quadrupolar GW mode, $(\ell,m)=(2,2)$, the GW phase is related to the orbital phase through
\begin{equation}
\Phi_{\rm GW}=2\Phi.
\end{equation}
Consequently, the instantaneous GW frequency is
\begin{equation}
f_{\rm GW}=
\frac{1}{2\pi}
\frac{d\Phi_{\rm GW}}{dt}=
\frac{1}{\pi}
\frac{d\Phi}{dt}.
\end{equation}

 The GW signal received by a distant observer is expressed in terms of the retarded time, defined as
\begin{equation}
t_{\rm ret}=t-r_*,
\end{equation}
where $t$ denotes the coordinate time at the GW source and $r_*$ is the tortoise coordinate. The retarded time accounts for the finite propagation time of the gravitational waves from the source to the location at which the waveform is extracted \cite{2021MNRAS.506.1007Z}. The tortoise coordinate is given by
\begin{equation}
r_*
=
r_A
+
2M_{\rm tot}
\ln\left(
\frac{r_A}{2M_{\rm tot}}-1
\right),
\end{equation}
with $r_A=\sqrt{\frac{A}{4\pi}},$ where $A$ denotes the proper surface area of the extraction sphere.

 For equal-mass binaries, the corresponding relative tidal correction term during the inspiral is given by \cite{2011PhRvD..83h4051V, PhysRevD.87.044001}
\begin{equation}
F_{\rm Tidal}(x)=\frac{52k_2}{5M_{\rm tot}C^5}x^5
\left(1+\frac{5203}{4368}x\right).
\end{equation}

The Taylor-T4 evolution consequently provides the time-dependent PN parameter $x(t)$ and orbital phase $\Phi(t)$, from which the GW frequency and phase are obtained. The dominant $(2,2)$ mode of the waveform is then used to determine the GW strain amplitude. In this way, the variation of the $\Delta$-meson coupling is connected systematically to the EOS, tidal deformability, and finally to the GW frequency, phase, and strain amplitude.
We emphasize that the TaylorT4 approximant provides a perturbative description of the inspiral and is not intended to model the nonlinear merger dynamics. 
The EOS-dependent differences discussed below should therefore be interpreted as late-inspiral trends within the adopted PN framework. In particular, quantitative predictions close to merger would require waveform models calibrated to numerical-relativity simulations and incorporating higher-order tidal contributions.

\section{Results}
\label{results}
\begin{figure*}
    \centering
    \includegraphics[angle=-90, width=15cm]{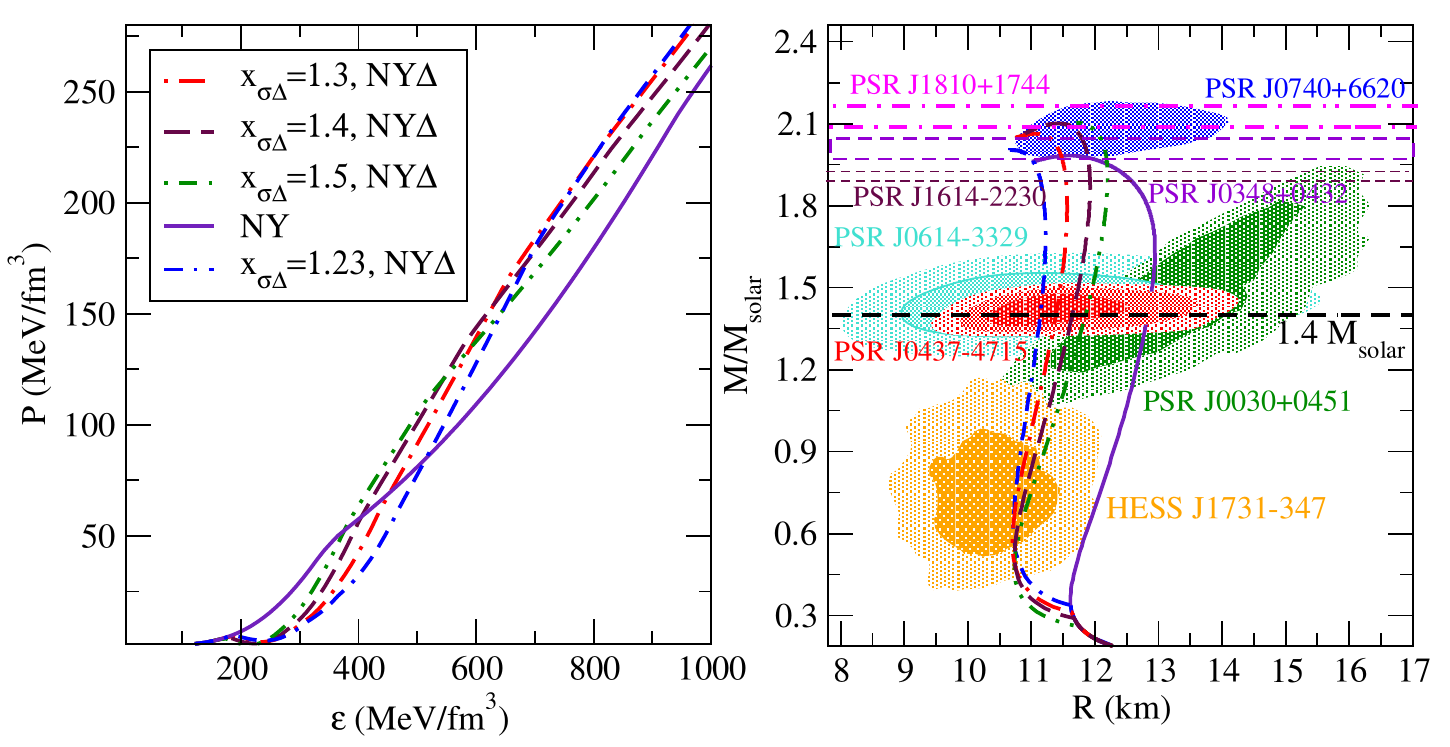}
    \caption{\textbf{Left panel:} Pressure as a function of energy density for the NY reference and representative NY$\Delta$ EOSs. \textbf{Right panel:} M-R relations for NY$\Delta$ matter with different $\Delta$--$\sigma$ potentials and for matter without $\Delta$ baryons. The \textcolor{black}{red, turquoise, blue, green and orange contours} represent the M-R confidence regions inferred for PSR J0437-4715\cite{choudhury_2024_13766753}, PSR J0614-3329 \cite{miller_2026_22131748}, PSR J0740+6620 \cite{Miller_2021, Riley_2021, Cromartie_2019}, PSR J0030+0451 \cite{Miller_2019, Riley_2019}, \textcolor{black}{and HESS J1731-347 \cite{2024PhRvD.109d3036H} respectively}. \textcolor{black}{The lighter, medium-darker, and darker shaded regions correspond to the \(99\%\), \(95\%\) and \(68\%\) credible regions, respectively.} Other observational constraints include PSR J1810+1744 \cite{Romani_2021}, PSR J1614-2230 \cite{Demorest_2010, Arzoumanian_2018, 2010ApJ...724L.199O}, and PSR J0348+0432 \cite{Antoniadis_2013, 2018ChJPh..56..292H, Bandyopadhyay:2021gnp}.}
    \label{fig:1}  
\end{figure*}

In our analysis, the EOS is initially obtained for different values of the scalar $\Delta$-meson coupling ratio $x_{\sigma\Delta}$. The corresponding $x_{\omega\Delta}$ values are varied consistently with the constraint on $x_{\sigma\Delta}-x_{\omega\Delta}$. Then for each coupling prescription, the stellar structure equations are solved to obtain $M$, $R$, $k_2$, and $\Lambda$.  

For different coupling parameters, the matter EOS and the corresponding M-R relations are shown in the left and right panels, respectively, of Fig. \ref{fig:1}. 
\textcolor{black}{The reference EOS with NY composition supports a maximum mass of \(1.9828\,M_\odot\). This is \(0.0972\,M_\odot\) ($\simeq1.39\sigma$) below the central mass estimate \(2.08\pm0.07\,M_\odot\) for PSR J0740+6620, corresponding to approximately \(1.39\) quoted standard deviations under a Gaussian approximation. The discrepancy represents mild tension rather than exclusion at high significance. We retain the NY EOS as a controlled reference with the same underlying interaction prescription and without \(\Delta\) baryons. The \(x_{\sigma\Delta}=1.23\) model supports \(2.0031\,M_\odot\), slightly below the lower edge of the quoted one-standard-deviation interval, while the other sampled NY\(\Delta\) models exceed that lower edge. 
Relative to the NY reference, the inclusion of \(\Delta\) baryons modifies the pressure in a density-dependent manner. Along the adopted coupling sequence, both the attractive scalar interaction and the repulsive vector interaction change, while charge neutrality and \(\beta\) equilibrium redistribute the particle populations. The resulting EOS behavior therefore reflects the combined response of the meson fields and composition. It cannot be attributed to the scalar coupling or to an increase in the number of particle species alone. The corresponding maximum masses generally increase over the sampled sequence, reaching their largest tabulated value at \(x_{\sigma\Delta}=1.45\), followed by a small decrease at \(x_{\sigma\Delta}=1.50\).}

The systematic change across the adopted $\Delta$ coupling sequence reflects the competition between the attractive scalar and repulsive vector interactions, both of which increase as $x_{\sigma\Delta}$ and $x_{\omega\Delta}$ are varied simultaneously. A larger scalar coupling reduces the effective $\Delta$ mass and generally favors an earlier appearance of $\Delta$ resonances, whereas the accompanying increase of the vector coupling enhances the repulsive interaction at higher density. The resulting EOS therefore exhibits a nontrivial density dependence determined by the evolving particle fractions rather than by the scalar interaction alone.
Consequently, the EOS becomes relatively stiffer in the intermediate-density region. At still higher densities, however, the larger $\Delta$ population leads to a stronger reduction in the Fermi pressure, causing the EOS to soften relative to models with smaller $x_{\sigma\Delta}$. These changes are reflected in the corresponding M-R relations, where increasing $x_{\sigma\Delta}$ results in larger maximum neutron-star masses. The threshold densities of $\Delta$ baryons, the maximum attainable mass and the tidal deformability of a $1.4~M_\odot$ star for different $x_{\sigma\Delta}$ are given in Table \ref{tab:delta_coupling_threshold}.

The negatively charged $\Delta^-$ is the first member of the $\Delta$ quartet to appear, with its threshold decreasing from $1.27n_0$ to $1.20n_0$ across the adopted coupling sequence.
The onset and subsequent population of the other $\Delta$ states exhibit a more non-monotonic behavior because of their competition with nucleons and hyperons under the simultaneous requirements of $\beta-$equilibrium and charge neutrality. 
In particular, the delayed onset of the $\Delta^{+}$ for the strongest coupling prescription and the absence of the $\Delta^{++}$ indicate that the evolution of the charged baryonic components cannot be understood solely in terms of the scalar interaction.

\begin{figure*}   
    \centering
    \includegraphics[angle=-90, width=15cm]{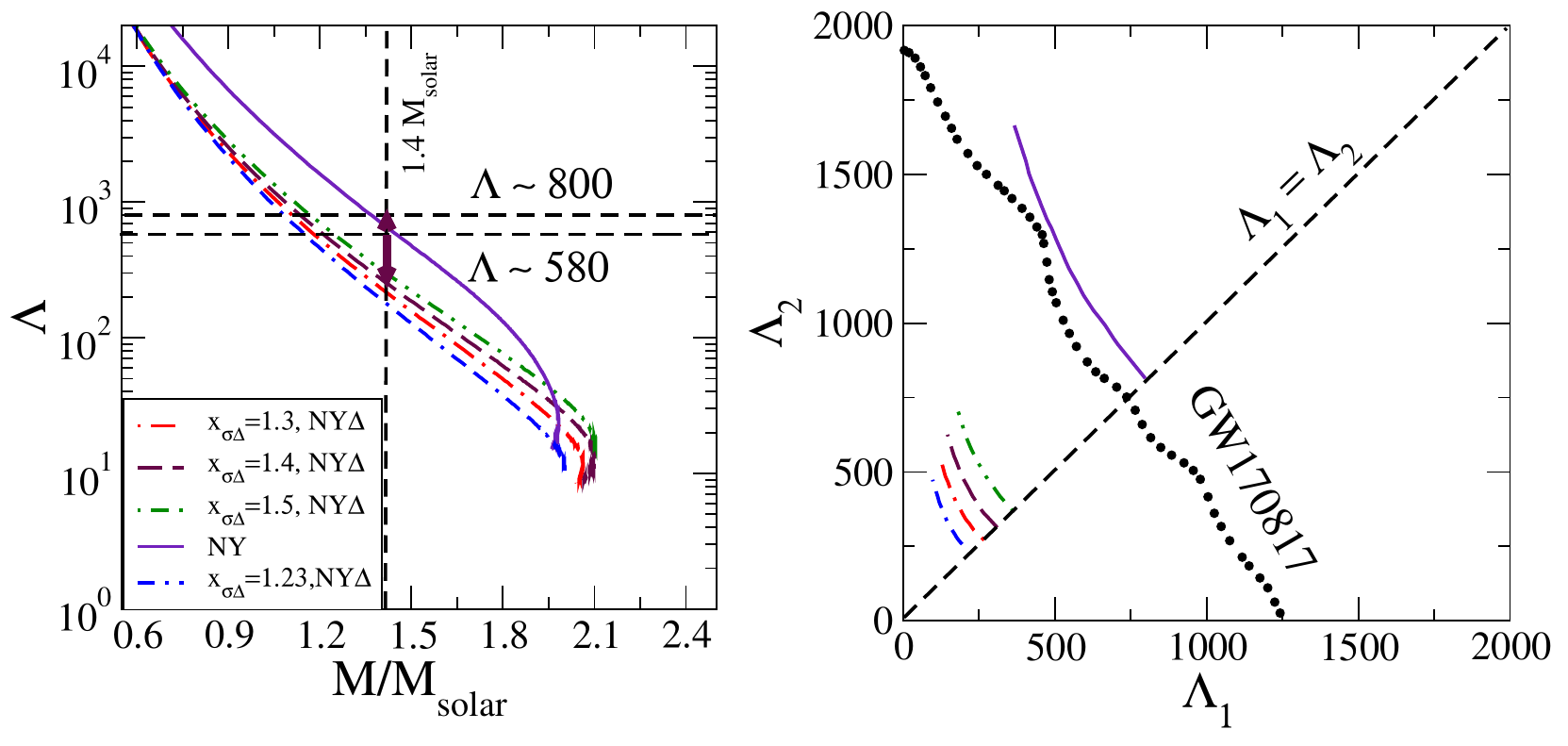}
   \caption{\textbf{Left panel:} Variation of dimensionless tidal deformability $\Lambda$ with NS mass $M$ for NY$\Delta$ matter and matter without $\Delta$ baryons. \textbf{Right panel:} contour of mutual tidal deformability $\tilde{\Lambda}$ in the plane of tidal deformabilities $\Lambda_1$ and $\Lambda_2$ of the two binary components for GW170817 \cite{PhysRevLett.119.161101, Abbott_2017b, Abbott_2017c}, considering NY$\Delta$ matter and matter without $\Delta$ baryons.}
    \label{fig:2}
\end{figure*}

\begin{table*}
\centering
\large
\setlength{\tabcolsep}{8pt}
\caption{Threshold densities of the $\Delta$ resonances in unit of $n_0 = 0.152~fm^{-3}$, maximum gravitational mass $M_{\mathrm{max}}$, and tidal deformability $\Lambda_{1.4}$ of a canonical NS for different values of the $\Delta$-scalar coupling $x_{\sigma\Delta}$.}
\label{tab:delta_coupling_threshold}
\resizebox{0.48\textwidth}{!}{%
\begin{tabular}{ccccccc}
\toprule
$x_{\sigma\Delta}$ & $x^{\Delta^{-}}_{th}$ & $x^{\Delta^{0}}_{th}$ & $x^{\Delta^{+}}_{th}$ & $x^{\Delta^{++}}_{th}$ & $M_{\text{max}}/M_\odot$ & $\Lambda_{1.4}$ \\
\midrule
No $\Delta$ & - & - & - & - & 1.9828 &691.8034 \\
1.23 & 1.27 & 2.11 & 2.46 & 2.74 & 2.0031 &187.9488 \\
\textcolor{black}{1.25} & \textcolor{black}{1.27} & \textcolor{black}{2.11} & \textcolor{black}{2.46} & \textcolor{black}{2.81} & \textcolor{black}{2.0232} & \textcolor{black}{203.4006}\\
1.3 & 1.27 & 2.04 & 2.46 & 2.95 & 2.0627 &231.0684 \\
\textcolor{black}{1.35} & \textcolor{black}{1.27} & \textcolor{black}{1.97} & \textcolor{black}{2.46} & \textcolor{black}{4.28} & \textcolor{black}{2.0873} & \textcolor{black}{265.9368}\\
1.4 & 1.27 & 1.97 & 2.60 & 4.77 & 2.1004 &261.8674 \\
\textcolor{black}{1.45} & \textcolor{black}{1.20} & \textcolor{black}{1.90} & \textcolor{black}{3.79} & \textcolor{black}{5.61} & \textcolor{black}{2.1049} & \textcolor{black}{287.9885}\\
1.5 & 1.20 & 1.83 & 4.63 & - &  2.1034 &310.2464 \\
\bottomrule
\end{tabular}}
\end{table*}

The variation of $\Lambda$ with the mass of the star is shown in the left panel of Fig. \ref{fig:2}. It is evident that with an increase of $x_{\sigma\Delta}$, the value of $\Lambda_{1.4}$ increases. GW observations, particularly those of GW170817, provide constraints on the tidal deformability of NSs, including an upper bound on $\Lambda$ for a $1.4~M_\odot$ star. The tidal deformability of a canonical $1.4~M_\odot$ NS predicted by the NY$\Delta$ EOSs is significantly lower than that obtained for the hypernuclear EOS without $\Delta$ baryons. Nevertheless, all NY$\Delta$ EOSs satisfy the updated constraint from the LIGO/Virgo Collaboration, which reports a value of $\Lambda_{1.4}=190^{+390}_{-120}$ at the $90\%$ credible level \cite{PhysRevLett.121.161101, PhysRevD.99.023009}. In the right panel of Fig.~\ref{fig:2}, we show the contours of the binary tidal deformability $\tilde{\Lambda}$ in the $\Lambda_1$--$\Lambda_2$ plane for the two components of the binary system. The contours corresponding to the $\Delta$-admixed EOS lie within the observationally allowed region, indicating that these models are consistent with the GW170817 tidal deformability constraints.

The resulting tidal deformabilities are incorporated into the PN description of the binary inspiral through the tidal correction, which subsequently generates the orbital evolution and waveform phase. For this calculation, we consider a canonical mass star of mass $1.4~M_\odot$. The radius ($R$), love number ($k_2$) and compactness ($C$) for a canonical star is presented in the Table~\ref{tab:delta_coupling}.

\begin{table}
\centering
\large
\caption{Properties of a $1.4~M_{\odot}$ NS for different values of the $\Delta$-scalar coupling $x_{\sigma\Delta}$ in hypernuclear matter admixed with $\Delta$ baryons.}
\label{tab:delta_coupling}
\resizebox{0.32\textwidth}{!}{%
\begin{tabular}{cccc}
\toprule
$x_{\sigma\Delta}$ & Radius (km) & $k_2$ & Compactness \\
\midrule
No $\Delta$ & 12.8341 & 0.1141 & 0.1615 \\
1.23 & 11.1432 & 0.0631 & 0.1862 \\
\textcolor{black}{1.25} & \textcolor{black}{11.2249} & \textcolor{black}{0.0647} & \textcolor{black}{0.1842}\\
1.3 & 11.3874 & 0.0683 & 0.1815 \\
\textcolor{black}{1.35} & \textcolor{black}{11.5089} & \textcolor{black}{0.0755} & \textcolor{black}{0.1801} \\
1.4 & 11.6505 & 0.0711 & 0.1784 \\
\textcolor{black}{1.45} & \textcolor{black}{11.7717} & \textcolor{black}{0.0739} & \textcolor{black}{0.1764} \\
1.5 & 11.8626 & 0.0766 & 0.1751 \\
\bottomrule
\end{tabular}}
\end{table}

\begin{figure}   
    \centering
    \includegraphics[angle=-90, width=8.5cm]{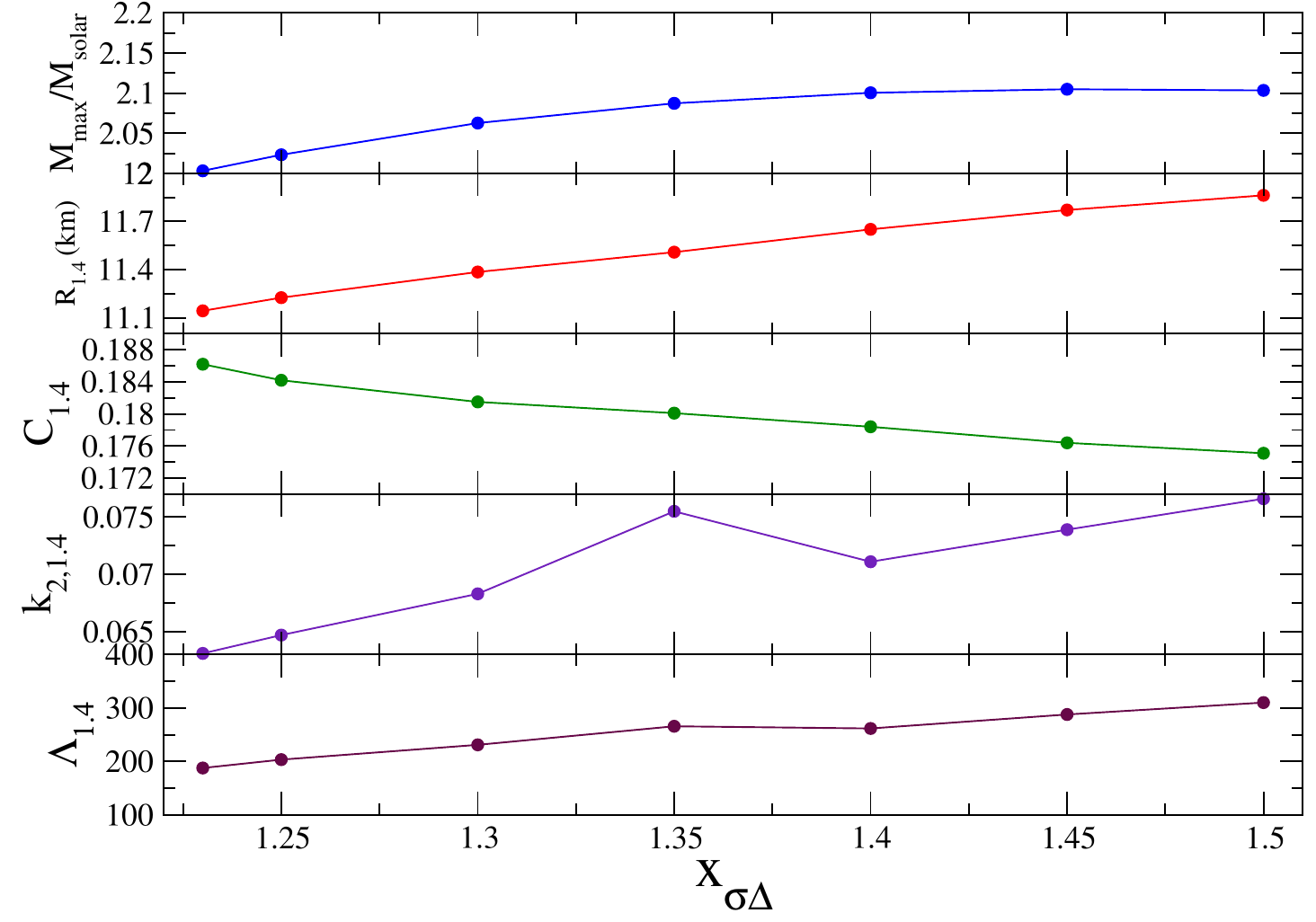}
    \caption{ Variation of the maximum mass $M_{\rm max}$, radius $R_{1.4}$, compactness $C_{1.4}$, tidal Love number $k_2$, and tidal deformability $\Lambda_{1.4}$ of NSs with the $\Delta$-meson coupling parameter $x_{\sigma\Delta}$ for the range $1.23 \leq x_{\sigma\Delta} \leq 1.5$ .}
    \label{fig:parameter_scan}
\end{figure}
\textcolor{black}{To systematically investigate the impact of the $\Delta$-resonance coupling on NS properties, we perform a scan over the parameter $x_{\sigma\Delta}$ and show the corresponding variations of the maximum mass, radius, compactness, Love number, and tidal deformability in Fig. \ref{fig:parameter_scan}. As $x_{\sigma\Delta}$ increases from 1.23 to 1.50, the maximum mass increases from $2.0031 M_\odot$ to $2.1034 M_\odot$, reaching a maximum value of $2.1049 M_\odot$ at $x_{\sigma\Delta}=1.45$. This indicates that the variation of the $\Delta$-meson coupling within the considered range can noticeably modify the high-density behavior of the EOS and consequently the maximum mass of the star. At $1.4 M_\odot$, the radius increases from $11.14$ km to $11.86$ km, while the compactness decreases from $0.1862$ to $0.1751$ over the same parameter range. The tidal Love number $k_2$ also changes with $x_{\sigma\Delta}$, increasing from $0.0631$ at $x_{\sigma\Delta}=1.23$ to $0.0766$ at $x_{\sigma\Delta}=1.50$, although a non-monotonic behavior is observed, particularly around $x_{\sigma\Delta}=1.40$. Since the tidal deformability depends sensitively on both the Love number and compactness through $\Lambda=(2/3)k_2C^{-5}$, $\Lambda_{1.4}$ generally increases from $187.95$ to $310.25$ across the scanned range. A small decrease is observed at $x_{\sigma\Delta}=1.40$, where $\Lambda_{1.4}$ decreases from $265.94$ at $x_{\sigma\Delta}=1.35$ to $261.87$, consistent with the corresponding decrease in $k_2$.}

\begin{figure}   
    \centering
    \includegraphics[angle=0, width=8.5cm]{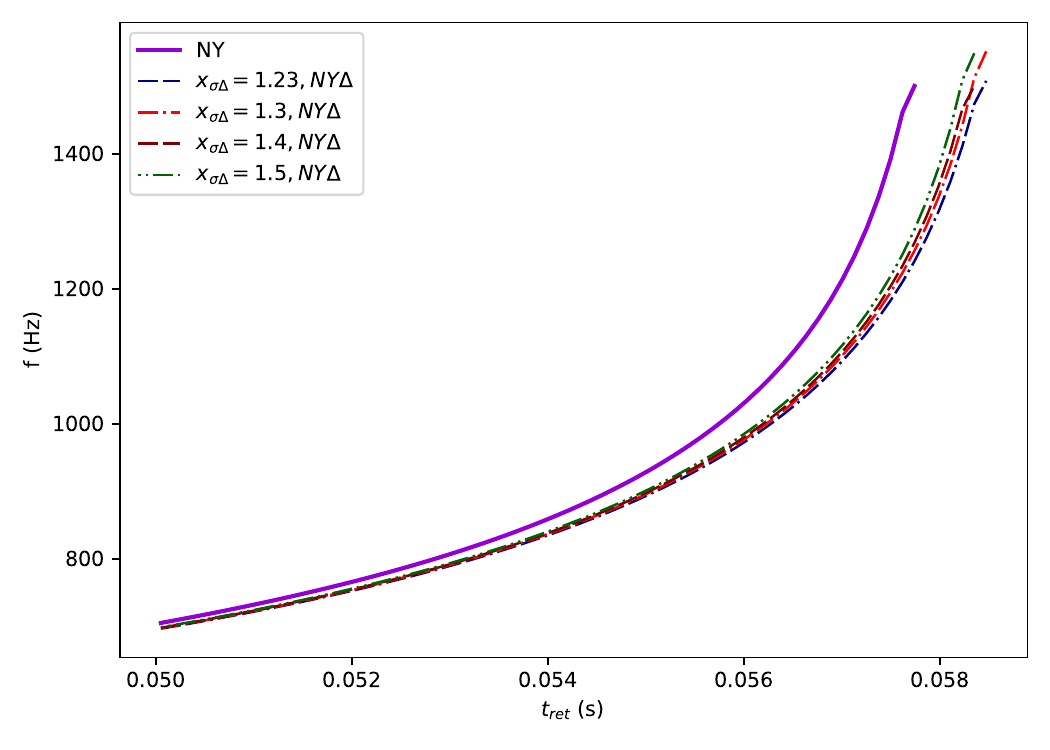}
    \caption{ Variation of GW frequency with retarded time for hyperonic matter with $\Delta$ baryons for different values of the $\Delta$--$\sigma$ coupling $x_{\sigma\Delta}$, and for hyperonic matter without $\Delta$ baryons.}
    \label{fig:freq_ret}
\end{figure}

Fig.~\ref{fig:freq_ret} shows the evolution of the GW frequency with the retarded time for equal-mass ($1.4+1.4~M_{\odot}$) binary neutron star systems described by the different EOSs considered in this work. The frequency increases monotonically with the retarded time for all EOSs, exhibiting the characteristic chirp behaviour expected during the inspiral phase as the binary loses orbital energy and angular momentum through the emission of GWs. The inspiral tracks remain nearly indistinguishable at earlier retarded times ($t_{\rm ret}\lesssim0.053~\mathrm{s}$), indicating that finite-size tidal effects are negligible when the stellar separation is relatively large. As the binary approaches merger, however, the frequency evolution becomes increasingly sensitive to the underlying EOS, resulting in a gradual separation of the curves.

The hypernuclear EOS (NY), which possesses the largest tidal deformability as given in Table \ref{tab:delta_coupling_threshold}, exhibits the most rapid increase in the GW frequency during the late inspiral. Consequently, the NY curve reaches the merger frequency at an earlier retarded time, terminating at $t_{\rm ret}=\textcolor{black}{0.057877}~\mathrm{s}$ with a final frequency of $\textcolor{black}{1539.83}~\mathrm{Hz}$. In contrast, the EOSs containing $\Delta$ baryons extend to slightly larger retarded times, with terminal values ranging between $0.058409~\mathrm{s}$ and $0.058525~\mathrm{s}$. This behaviour reflects the modifications in the stellar structure introduced by the presence of $\Delta$ resonances, which alter the tidal response of the binary during the late stages of inspiral.

Within the NY$\Delta$ family, the influence of the $\Delta$-meson coupling strength is clearly evident. The EOS with $x_{\sigma\Delta}=1.23$, corresponding to the smallest tidal deformability ($\Lambda=187.94$), exhibits the lowest frequency evolution throughout the inspiral and reaches a final frequency of \textcolor{black}{$1483.63~\mathrm{Hz}$}. As the coupling strength increases, the frequency curves shift progressively upward. The EOS with \textcolor{black}{$x_{\sigma\Delta}=1.5$} attain the highest final frequency of \textcolor{black}{$1535.63~\mathrm{Hz}$}. 

Another feature is that the differences among the EOSs become appreciable only during the final stage of the inspiral, when the GW frequency exceeds approximately $1000~\mathrm{Hz}$. At lower frequencies, the orbital dynamics are dominated by the point-particle contribution, leading to nearly identical inspiral tracks. As the orbital separation decreases, tidal interactions become increasingly important, resulting in measurable differences in the frequency evolution. Although the variation in the final frequency is only of the order of a few tens of hertz, these differences are systematic and directly reflect the influence of the microscopic composition of dense matter. Such EOS-dependent signatures in the late inspiral waveform provide a promising avenue for constraining the presence of exotic particles, including $\Delta$ resonances, with current and future GW observations.

\begin{figure}   
    \centering
    \includegraphics[angle=0, width=8.5cm]{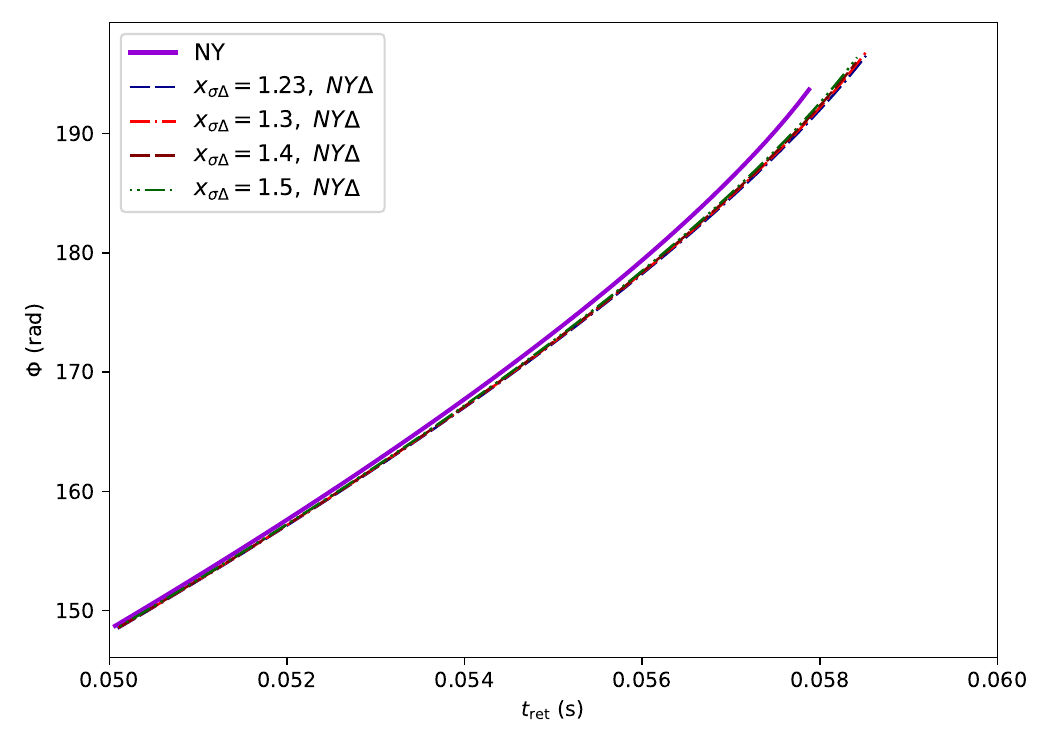}
\caption{ Variation of GW phase with retarded time for hyperonic matter with $\Delta$ baryons for different values of the $\Delta$--$\sigma$ coupling $x_{\sigma\Delta}$, and for hyperonic matter without $\Delta$ baryons.}
    \label{fig:phase}
\end{figure}

Fig.~\ref{fig:phase} presents the evolution of the GW phase with the retarded time for the different $x_{\sigma\Delta}$ cases. Similar to the frequency evolution, the phase increases monotonically throughout the inspiral, reflecting the continuous accumulation of orbital cycles as the binary evolves toward merger. {\color{black}{During the early inspiral ($t_{\rm ret}\lesssim0.053$ s), the phase evolution of the different EOS models remains relatively similar. As the binary evolves toward the late inspiral, differences in the phase evolution become more apparent, reflecting the different frequency evolution associated with the underlying EOSs.}

\begin{figure}   
    \centering
    \includegraphics[angle=0, width=8.5cm]{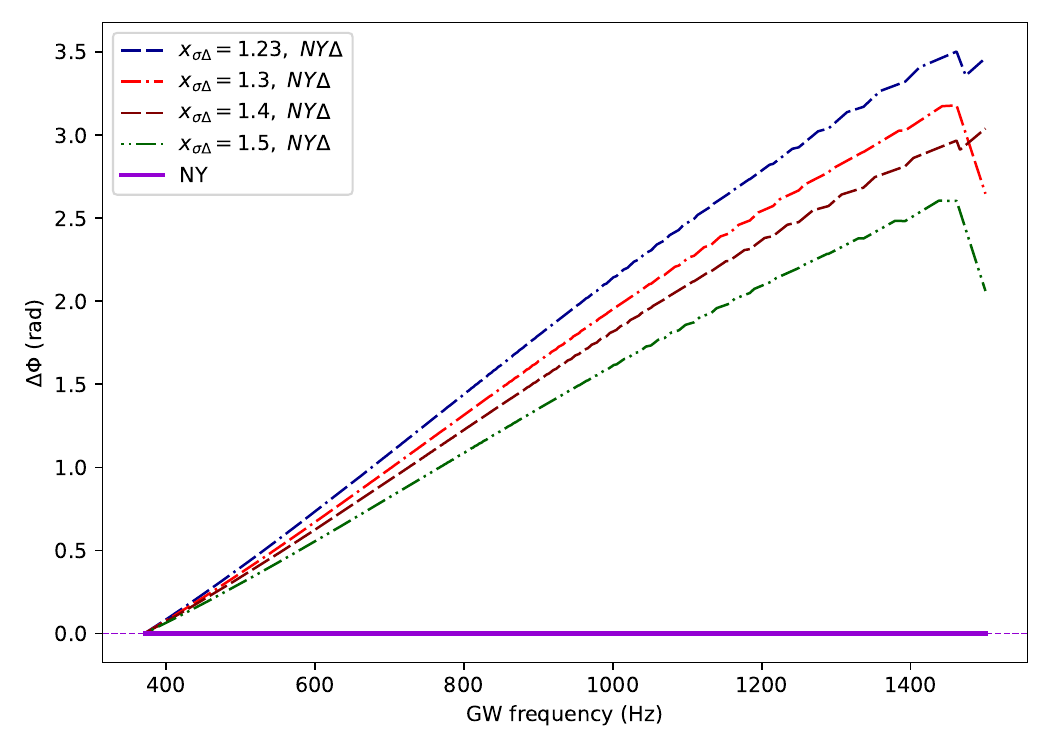}
    \caption{\textcolor{black}{Accumulated GW phase difference relative to the NY reference as a function of GW frequency. All waveforms are phase-aligned at \(f_{\rm ref}=371\) Hz and evaluated on the same frequency grid. Positive values indicate greater accumulated phase for the NY\(\Delta\) waveform between the reference and comparison frequencies.}}
    \label{fig:phase_freq}
\end{figure}

The relation between the GW phase and frequency is given by
\begin{equation}
    f(t_{\rm ret})=\frac{1}{2\pi}\frac{d\Phi}{dt_{\rm ret}}.
\end{equation}
Thus, the phase evolution represents the cumulative effect of the instantaneous frequency evolution. In Fig. \ref{fig:freq_ret}, the NY model develops a more rapid increase in GW frequency during the late inspiral compared with the NY$\Delta$ models. This is reflected in Fig. \ref{fig:phase}, where the phase accumulates at a rate determined by the instantaneous frequency. The phase curves corresponding to the different $x_{\sigma\Delta}$ values remain comparatively close to one another, indicating that their differences in phase accumulation are smaller than the difference between the NY and NY$\Delta$ families over most of the inspiral. For a quantitative comparison of the EOS-dependent phase evolution, we align the waveforms at the common reference frequency \(f_{\rm ref}=371\) Hz and evaluate the phase difference at the same GW frequency. The resulting phase evolution as a function of frequency is shown in Fig. \ref{fig:phase_freq}. At a common GW frequency of \(f=1400\) Hz, the phase differences between the NY and NY$\Delta$ models are approximately 3.36, 3.05, 2.85, and 2.50 rad for \(x_{\sigma\Delta}=1.23\), 1.30, 1.40, and 1.50, respectively.}

\begin{figure*}   
    \centering
    \includegraphics[angle=0, width=11cm, height=6.5cm]{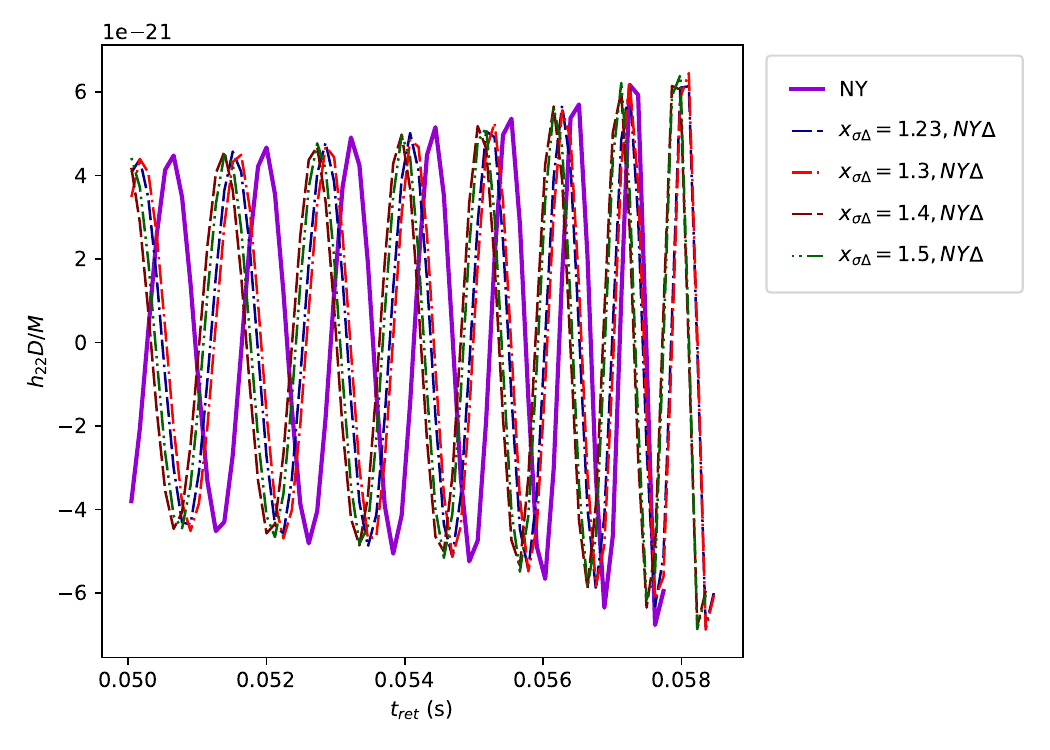}
    \caption{ Variation of GW phase with frequency for hyperonic matter with $\Delta$ baryons for different values of the $\Delta$--$\sigma$ coupling $x_{\sigma\Delta}$, and for hyperonic matter without $\Delta$ baryons.}
    \label{fig:strain}
\end{figure*}

\begin{table*}
\centering
\large
\setlength{\tabcolsep}{8pt}
\caption{Final values of the retarded time, gravitational-wave frequency, phase, \textcolor{black}{phase difference with respect to NY model at 1400 Hz} and scaled strain amplitude for different values of $x_{\sigma\Delta}$, and for hyperonic matter without $\Delta$ baryons.}
\label{tab:waveform_final}
\resizebox{0.62\textwidth}{!}{%
\begin{tabular}{lccccc}
\hline
$x_{\sigma\Delta}$ & $t_{\rm ret}$ (s) & $f$ (Hz) & $\Phi$ (rad) &  \textcolor{black}{$\Delta \Phi$ rad (at 1400 Hz)}  & $|h_{22}D/M|\times10^{-21}$ \\
\hline
No $\Delta$ & \textcolor{black}{0.057877} & \textcolor{black}{1539.827} & \textcolor{black}{193.681} & \textcolor{black}{- }& \textcolor{black}{8.0128} \\
1.23 & \textcolor{black}{0.058518} & \textcolor{black}{1483.632} & \textcolor{black}{196.508} &\textcolor{black}{3.36}& \textcolor{black}{7.7536} \\
1.3 & \textcolor{black}{0.058525} & \textcolor{black}{1528.840} & \textcolor{black}{196.882} &\textcolor{black}{3.05}& \textcolor{black}{7.9542} \\
1.4 & \textcolor{black}{0.058409} & \textcolor{black}{1484.096} & \textcolor{black}{195.964} &\textcolor{black}{2.85}& \textcolor{black}{7.7567} \\
1.5 & \textcolor{black}{0.058418} & \textcolor{black}{1535.626} & \textcolor{black}{196.378} &\textcolor{black}{2.50}& \textcolor{black}{7.9861} \\
\hline
\end{tabular}}
\end{table*}
Figure \ref{fig:strain} shows the real part of the dominant quadrupolar mode as a function of the aligned retarded time. Its amplitude envelope grows as the orbital frequency increases. The amplitude expression used here contains point-particle corrections but no explicit tidal amplitude terms. Consequently, at fixed component masses, distance, and frequency, the amplitude envelope is independent of the EOS. Differences visible at the same retarded time arise through the EOS-dependent frequency and phase evolution. Likewise, amplitude values evaluated at unequal terminal frequencies do not provide an independent diagnostic of the microscopic composition.

The final values of the scaled strain amplitude remain remarkably similar for all EOSs, varying only from \textcolor{black}{$7.75\times10^{-21}$ to $8.01\times10^{-21}$}. \textcolor{black}{ Among the NY$\Delta$ models, the configuration with $x_{\sigma\Delta}=1.23$ yields the smallest amplitude, $7.754\times10^{-21}$, closely followed by the $x_{\sigma\Delta}=1.4$ model, with $7.757\times10^{-21}$. The amplitudes increase for $x_{\sigma\Delta}=1.3$ and $1.5$, reaching $7.954\times10^{-21}$ and $7.986\times10^{-21}$, respectively. The NY model, without $\Delta$ baryons, exhibits the largest terminal amplitude among the considered EOSs. This non-monotonic variation indicates that the terminal GW strain amplitude depends on the adopted $\Delta$ meson coupling and the corresponding neutron-star properties. The nearly identical amplitudes for $x_{\sigma\Delta}=1.23$ and $1.4$, followed by higher values for $x_{\sigma\Delta}=1.3$ and $1.5$, suggest that the amplitude does not vary systematically with the coupling parameter. Instead, its response reflects the combined influence of the EOS-dependent stellar structure and tidal properties on the inspiral waveform. Nevertheless, the relatively small amplitude variation across these models suggests that the terminal strain amplitude alone may not provide a sufficiently sensitive discriminator of the different NY$\Delta$ EOSs, making the frequency and phase evolution potentially more informative for distinguishing their gravitational-wave signatures.}

Taken together, the frequency, phase, and strain evolution demonstrate that the inclusion of $\Delta$ resonances produces measurable signatures in the late-inspiral gravitational waveform. While the amplitude exhibits only modest EOS dependence, the accumulated phase and the frequency evolution show significantly greater sensitivity to the microscopic composition of dense matter. The final values of retarded time, GW frequency and scaled strain amplitude have been shown in Table \ref{tab:waveform_final}.\\

{\color{black}To quantify the distinguishability of the gravitational-wave signals predicted by the NY and NY$\Delta$ models, we calculate the noise-weighted waveform mismatch \cite{1996PhRvD..53.6749O} using the design-sensitivity noise power spectral densities (PSDs) of Advanced LIGO Plus (A+), the Einstein Telescope (ET), and Cosmic Explorer (CE). The comparison is performed over a common frequency interval of \(371\)–\(1400\) Hz, corresponding to the late-inspiral portion of the signals considered in this work. We use the plus-polarization waveforms and maximize the normalized overlap over relative time and phase shifts. The mismatch is defined as
\begin{equation}
    \mathcal{M}=1-\mathcal{O},
\end{equation}
where \(\mathcal{O}\) denotes the maximized, noise-weighted normalized overlap between the two waveforms.

As a characteristic estimate of the signal-to-noise ratio (SNR) associated with waveform distinguishability, we employ \cite{2008PhRvD..78l4020L}
\begin{equation}
    \rho_{\rm crit}=\frac{1}{\sqrt{2\mathcal{M}}}.
\end{equation}
The quantity \(\rho_{\rm crit}\) is a characteristic separation scale obtained from the small-difference relation \((\delta h|\delta h)\simeq2\rho^2\mathcal M\) for equal-norm waveforms after the stated alignment. Here, \(\rho\) refers to the SNR accumulated within the same 371–1400 Hz interval used to calculate the mismatch. It must not be identified with the full-band detection SNR. This estimate is not a confidence-level threshold for identifying \(\Delta\) baryons and does not account for degeneracies with masses, spins, or alternative EOS prescriptions.
The resulting mismatches and characteristic critical SNRs are presented in Table \ref{tab:mismatch_snr}. For the considered models, the mismatch ranges from \(3.98\times10^{-3}\) to \(7.19\times10^{-3}\), with corresponding characteristic critical SNRs ranging from approximately \(11.20\) to \(8.33\). The results indicate that the differences between the NY and NY\(\Delta\) waveforms are relatively small after maximizing the overlap over time and phase. The \(x_{\sigma\Delta}=1.4\) and \(1.5\) models exhibit smaller mismatches and consequently higher characteristic critical SNRs than the \(x_{\sigma\Delta}=1.23\) and \(1.3\) models across all three detector noise curves.

These estimates provide a quantitative measure of the waveform differences in the selected frequency band and indicate the characteristic SNR scale at which such differences may become relevant. However, the critical SNR derived from the mismatch is not the actual SNR of a source or a direct measure of its detection probability. A definitive assessment of EOS distinguishability would require incorporating the source-dependent SNR, parameter-estimation uncertainties, and waveform-model systematics.

\begin{table}[h]
\centering
{\color{black}
\caption{Noise-weighted waveform mismatch $\mathcal{M}$ and characteristic
critical SNR $\rho_{\rm crit}$ for comparisons between the NY and NY$\Delta$
models over the frequency interval $371$--$1400$ Hz, using
the design-sensitivity PSDs of A+, ET, and CE.}
\label{tab:mismatch_snr}
\begin{tabular}{lccc}
\hline
Detector & $x_{\sigma\Delta}$ & \quad $\mathcal{M} \times10^{-3}$
& $\rho_{\rm crit}$ \\
\hline
A+ & 1.23 & \quad $7.1989$ & 8.3340 \\
A+ & 1.30 & \quad $6.4567$ & 8.7999 \\
A+ & 1.40 & \quad $5.3261$ & 9.6890 \\
A+ & 1.50 & \quad $4.4810$ & 10.5632 \\
\hline
ET & 1.23 & \quad $6.6068$ & 8.6993 \\
ET & 1.30 & \quad $5.9594$ & 9.1597 \\
ET & 1.40 & \quad $4.9743$ & 10.0258 \\
ET & 1.50 & \quad $4.2222$ & 10.8816 \\
\hline
CE & 1.23 & \quad $6.2855$ & 8.9189 \\
CE & 1.30 & \quad $5.6809$ & 9.3816 \\
CE & 1.40 & \quad $4.6783$ & 10.3380 \\
CE & 1.50 & \quad $3.9816$ & 11.2060 \\
\hline
\end{tabular}}
\end{table}}

\section{Summary and Conclusions}
\label{Conclusions}

In this work, we investigated the influence of $\Delta$ baryons on the inspiral GW signal from equal-mass binary NS systems within the framework of the DDRMF model. The NS matter was composed of nucleons, hyperons, leptons, and the $\Delta$ resonances, with the $\Delta$-meson coupling strength varied over the range $x_{\sigma\Delta}=1.23$--$1.5$. For each EOS, the stellar structure was determined by solving the TOV equations, from which $k_2$, C, and $\Lambda$ were calculated. These quantities were subsequently incorporated into the TaylorT4 waveform approximant to examine the impact of the EOS on the inspiral GW observables. The inclusion of $\Delta$ baryons significantly modifies the tidal properties of NSs through changes in the EOS. Compared with the hypernuclear dense matter EOS, the NY$\Delta$ EOSs exhibit considerably smaller tidal deformabilities, reflecting the increased compactness of the stellar configurations. \textcolor{black}{These EOS-induced changes in the stellar structure subsequently affect the GW signal, primarily through modifications of the compactness and tidal deformability, rather than through a unique direct signature of $\Delta$ resonances. As a result, the presence of $\Delta$ baryons can leave imprints on the late-inspiral GW signal, particularly in its frequency and phase evolution.} The GW frequency increases monotonically with the retarded time for all EOSs, exhibiting the characteristic chirp behaviour expected during binary inspiral. While the inspiral tracks remain nearly identical at low frequencies, modeled EOS-dependent deviations develop during the final stage of the evolution where tidal interactions become increasingly important. 
\textcolor{black}{After alignment at the common reference frequency, the hypernuclear EOS reaches a specified higher GW frequency earlier than the NY$\Delta$ EOSs; the integration endpoints are not interpreted as physical merger frequencies.}
These variations arise from the distinct tidal responses associated with the different microscopic compositions of dense matter.

\textcolor{black}{The accumulated GW phase is sensitive to the presence of \(\Delta\) baryons, particularly during the late inspiral, where differences in the tidal properties of the EOSs become increasingly significant. To ensure a consistent comparison, the waveforms are aligned at a common reference frequency of \(f_{\rm ref}=371\) Hz, and their phase difference is evaluated at the same GW frequency. At \(f=1400\) Hz, the phase differences between the NY and NY\(\Delta\) models are approximately \(3.36\), \(3.05\), \(2.85\), and \(2.50\) rad for \(x_{\sigma\Delta}=1.23\), \(1.30\), \(1.40\), and \(1.50\), respectively. The decrease in the phase difference with increasing \(x_{\sigma\Delta}\) reflects the sensitivity of the accumulated phase to the EOS-dependent tidal properties and frequency evolution. Such phase differences are relevant to gravitational-wave data analysis because matched-filtering techniques are sensitive to the phase evolution of the waveform.} The evolution of the dominant $(2,2)$ GW mode demonstrates that the waveform amplitude is comparatively less sensitive to EOS variations than either the frequency or the accumulated phase. 
\textcolor{black}{The adopted amplitude expression has no explicit tidal correction, so differences evaluated at unequal terminal frequencies should not be interpreted as independent EOS signatures.}
This behaviour indicates that the principal imprint of the EOS on the inspiral waveform is encoded in its timing and phase evolution rather than in substantial modifications of the waveform amplitude. 
\textcolor{black}{The mismatch estimates suggest characteristic critical SNRs of approximately 8–11 within the analyzed 371–1400 Hz band, with masses and spins held fixed; these are preliminary waveform-separation estimates rather than composition-identification thresholds. Overall, $\Delta$ baryons modify the modeled inspiral signal through their influence on tidal deformability, but these effects are not unique to their presence; any constraints on the $\Delta-$meson couplings remain conditional on the adopted EOS framework.}
While the amplitude differences remain relatively modest, the frequency and, in particular, the accumulated phase exhibit modeled EOS-dependent variations that may \textcolor{black}{ help distinguish between the NY and NY$\Delta$ models. The mismatch analysis further provides a preliminary estimate of the SNR scale at which these waveform differences may become relevant for GW observations with current and next-generation detectors.} These results highlight the potential of precision GW astronomy as a powerful probe of exotic degrees of freedom in dense matter and motivate the development of waveform models incorporating realistic EOS containing hyperons and $\Delta$ resonances. Future investigations including higher-order tidal effects, numerical-relativity waveforms, and Bayesian parameter estimation will further improve our understanding of the role of exotic matter in NS mergers.

\section*{Acknowledgements}
\textcolor{black}{The authors thank the anonymous referee for the valuable suggestions provided, which have greatly contributed to enhancing the quality of this manuscript.}
M.S. and A.S. acknowledge the financial support from the Science and Engineering Research Board (SERB), Department of Science and Technology, Government of India, through Project No. CRG/2022/000069. M.S. acknowledges partial financial support from the DRDO through Project No. DGTM/ERIP/GIA/24-25/010/005.
V.B.T. acknowledges the financial support from the Indian Space Research Organization (ISRO), Department of Space, Government of India, through Project No. RAC-S/GU/2024/4/22.

\bibliographystyle{unsrt}
\bibliography{references}
\end{document}